# Optical evidence for blue shift in topological insulator bismuth selenide in the few-layer limit

Yub Raj Sapkota[1], Asma Alkabsh[1], Aaron Walber[1], Hassana Samassekou[1], and Dipanjan Mazumdar[1,2]

1. Department of Physics, Southern Illinois University, Carbondale, IL 62901

2. Materials Technology Center, Southern Illinois University, Carbondale, IL, 62901

## ABSTRACT

Optical band gap properties of high-quality few-layer topological insulator $Bi_2Se_3$ thin films grown with magnetron sputtering are investigated using broadband absorption spectroscopy. We provide direct optical evidence of a rigid blue-shift to up to 0.5 eV in the band gap of $Bi_2Se_3$ as it approaches the two-dimensional limit. The onset of this behavior is most significant below six quintuple layers. The blue shift is very robust and is observed in both protected (capped) and exposed (uncapped) thin films. Our results are consistent with observations that finite-size effects have profound impact on the electronic character of topological insulators, particularly when the top and bottom surface states are coupled. Our result provides new insights, and the need for deeper investigations, into the scaling behavior of topological materials before they can have significant impact on electronic applications.

# INTRODUCTION

Topological insulators (TI), such as $Bi_2Se_3$ have gained much attention in basic and applied physics research due to the existence of "topologically-protected" gapless surface states [1, 2, 3]. These novel quantum states are robust in nature as they are immune to non-magnetic impurities as the electron momentum is locked to its spin [4]. Combined with the observation that these states can be sustained at room temperature in real materials with strong spin-orbit coupling (such as BiSb, $Bi_2Se_3$, $Sb_2Te_3$) and without the application of an external magnetic field have made them very interesting for device applications. Therefore, much of current research is focused on investigating the consequences of the protected metallic states, and it is envisioned that they are suited for advanced applications in areas such as spintronics and fault-tolerant quantum computing [5, 6, 7, 8, 9, 10].

$Bi_2Se_3$ is a prototypical example among 3D Topological insulators [4]. Single Dirac cone was observed at the $\Gamma$ point in bulk $Bi_2Se_3$ through angle-resolved photo emission (ARPES) [3, 11] [12] and scanning tunneling microscopy measurements [13]. It is classified as a strong topological insulator [14], where surface states retain zero-gap despite the presence of atomic-level non-magnetic impurities. The literature has also been growing regarding $Bi_2Se_3$ thin films. Recent experiments involving $Bi_2Se_3$ thin films demonstrated proximity-induced superconductivity [15, 16] and ferromagnetism [5], both phenomena associated with symmetry breaking. Exotic effects such as Quantum anomalous Hall effect has been experimentally observed in magnetic TI thin films at ultralow temperatures [17, 18, 19].

Many potential applications of TIs like $Bi_2Se_3$ will rely on their scaling behavior. It is therefore important and intriguing to ask as to what happens to such exotic materials as they approach the two-dimensional limit? Thin films, therefore, provide an ideal platform to investigate low dimensional physics of TIs. One of the more intriguing consequence of finite-size effects in TIs is the opening of an energy gap in the surface states due to quantum tunneling between the top and bottom surfaces, an effect which was first pointed out by theory [20, 21, 22], and also verified later through first-principles calculations [23] [24]. Experimentally such an effect was directly verified by Zhang *et al.* [25] and Sakamoto *et al*. [26] in ultra-thin $Bi_2Se_3$ films (below six quintuple layers) through angle resolved photo emission spectroscopy (ARPES) measurement. The observed gap opening is substantially large (~ few tenths of eV). Weak localization effects can also result in a gap opening in few-layer $Bi_2Se_3$ thin films, but it is of the order of meV [27, 28]. Also, recently, Vargas *et al.* reported large blue-shift in $Bi_2Se_3$ nanoparticles, which they attributed to quantum-confinement effects in all directions [29]. All these reports demonstrate that finite-size can have a profound impact in topological materials. In this work, we report that the optical band gap changes can also occur in such materials in the 2-dimensinal limit.

Investigations of optical properties of $Bi_2Se_3$ thin films has been a subject of previous studies. Variation of optical properties with thickness has also been noted. For example, Post *et al.* [30] studied uncapped 15-99 quintuple layers (QL) $Bi_2Se_3$ films and found band gap values below 0.3 eV that are attributed to impurity states or surface contamination. Eddrief *et al.* [31] measured optical properties of 3-54 QL $Bi_2Se_3$ thin films. While they cover a broad thickness range, the optical properties of the 3 QL film do show a behavior that is consistent, but not clearly reported, with an increase in band gap. Higher optical transmittance in 5 or 6 QL layer $Bi_2Se_3$ film has been reported that implies higher band gap [32, 33]. However, to the best of our knowledge, a systematic bandgap investigation is lacking at the two-dimensional (few-layer) limit of $Bi_2Se_3$.

Here we report on the optical properties of few-layer $Bi_2Se_3$ (~2-10 QL) thin films. High-quality $Bi_2Se_3$ films (both capped and uncapped) were grown on $Si/SiO_2$ and quartz substrates using radio frequency magnetron sputtering. We have discovered a clear increase in the bulk band gap of $Bi_2Se_3$ to the tune of 0.5 eV as the thickness is reduced from ~10 QL to 2 QL. Such an increase is consistent with drastic changes in the electronic properties at lower dimensions. The blue-shift is robust and observed in samples with and without a protective capping layer. Our work consolidates a growing number of studies that highlight finite-size effects in topological materials.

## EXPERIMENTAL DETAILS

In this work, high-quality $Bi_2Se_3$ few-layer thin films were fabricated in the 2-10 nm thickness range that roughly translates to 2-10 quintuple layers (1 QL ~ 0.95 nm). $Bi_2Se_3$ was grown using commercially available stoichiometric target and RF sputtered in a high vacuum magnetron sputtering system (base pressure $4 \times 10^{-9}$ Torr). $Bi_2Se_3$ films were grown at room temperature and annealed *in-situ* at 300 C. We recently employed a similar method to grow other large-area layered materials such as $MoS_2$ [34]. To protect the surface from contamination and oxidation, and yet retain optical transparency in the infrared and visible wavelengths, some $Bi_2Se_3$ thin films were capped with an amorphous BN layer grown *in-situ* at room temperature.

Structural and interface properties were characterized by means of high-resolution X-ray diffraction and reflectivity using a Rigaku Smartlab Diffractometer equipped with a 220-Ge monochromator to obtain a Cu K$\alpha_1$ radiation. Raman spectroscopy was employed to confirm the vibrational modes of $Bi_2Se_3$ using a Nanophoton Raman-11 with a 532 nm laser. The laser power was kept low at 10 mW to avoid local heating. Optical constants such as complex dielectric constant ($\varepsilon_1, \varepsilon_2$) were investigated using spectroscopic ellipsometry (JA Wollam M2000V, 1.1-3.0 eV), and optical band gap of few layer-$Bi_2Se_3$ films was measured using a broadband optical spectrometer (Shimadzu UV3600) in the 0.375-6.2 eV range.

**RESULTS AND DISCUSSION**

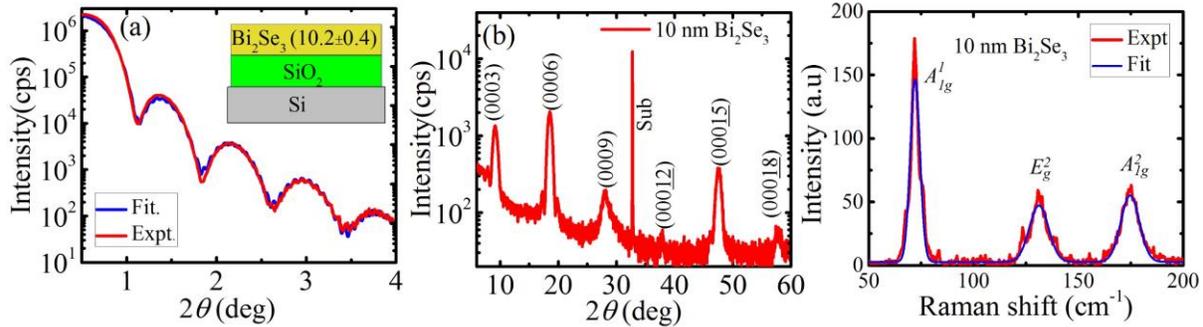

*Figure 1 (a) High-resolution x-ray reflectivity of ~ 10 nm $Bi_2Se_3$ thin film grown on $Si/SiO_2$ substrate. Inset shows the thickness and roughness value of $Bi_2Se_3$ film as obtained from the fit of reflectivity data. (b) X-ray diffraction pattern of 10 nm $Bi_2Se_3$ film showing only (000l) peaks, implying out of plane growth. (c) Raman vibration modes of the $Bi_2Se_3$ film showing the characteristic $A^1_{1g}$, $E^2_g$, $A^2_{1g}$ modes.*

Figure 1 shows the structural characteristics of an uncapped 10nm $Bi_2Se_3$ thin film grown on $Si/SiO_2$ substrate. In Fig. 1a we show the X-ray reflectivity data of the film. Oscillatory thickness pattern from the ~10 nm $Bi_2Se_3$ layer and the 100 nm $SiO_2$ layer are observed. This is indicative of sharp interfaces. The thickness and roughness value as obtained from the reflectivity fit [35, 36] is shown in the inset of Fig 1(a). Roughness of ~0.4 nm is less than half of quintuple layer. The extracted density of the $Bi_2Se_3$ film is also in very good agreement with bulk value (~8 × $10^{-3}$ g/cm$^3$).

Fig. 1b shows the high-resolution theta-2theta scan of the X-ray diffraction pattern of the 10 QL $Bi_2Se_3$ thin film. Clear diffraction peaks can be identified for the (002) silicon substrate (labeled as "sub") and (000l) peaks of the $Bi_2Se_3$ layer. This is indicative of out-of-plane growth. Thickness fringes are also observed around the (0003) and (0006) that are consistent with very smooth films. Off-axis measurements on high Miller indices peaks gave a=4.17 Å, c=28.56 Å. Taken together with the reflectivity data, we confirm that the properties of few-layer $Bi_2Se_3$ films are of superior bulk and interface quality and compares very favorably with Molecular Beam Epitaxy (MBE) grown films.

Additional crystalline structure characterization was also performed using Raman spectroscopy that has become very effective in investigating layered materials. The raw data and the line shape fits are shown in Fig 1c. Clear Raman modes were observed at ~72.0 cm$^{-1}$, 131.1 cm$^{-1}$ and 174.6 cm$^{-1}$ that correspond to $A^1_{1g}$, $E^2_g$ and $A^1_{2g}$ modes of $Bi_2Se_3$ [37]. The full width at half maxima is ~5 cm$^{-1}$ for the $A^1_{1g}$ mode, and ~9-10 cm$^{-1}$ for both the $E^2_g$ and $A^1_{2g}$ modes. These values are comparable to few-layer single crystal counterparts [37] . Raman data reiterates the high crystalline quality of the grown $Bi_2Se_3$ thin films.

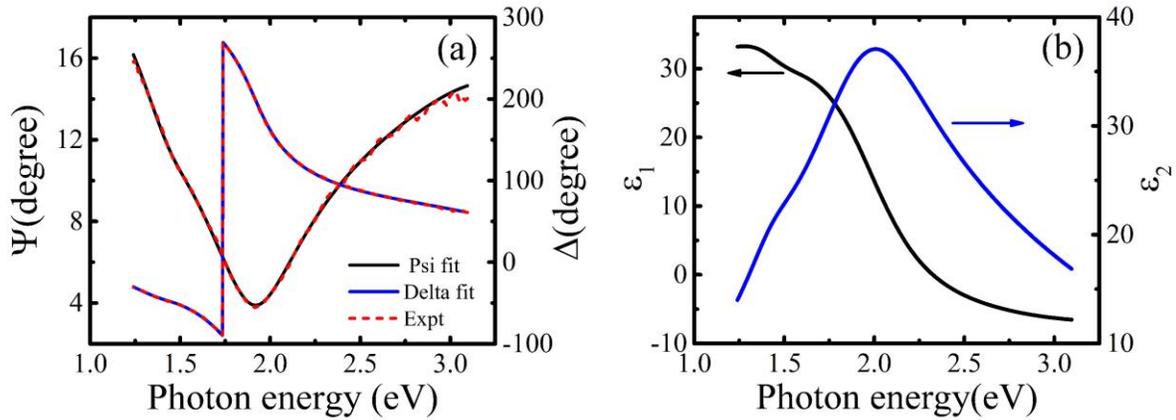

*Figure 2 (a) Experimental and fitted Psi ($\Psi$) and Delta ($\Delta$) for the 10 nm $Bi_2Se_3$ sample for incident angle of 70 degrees. (b) Extracted complex dielectric constants ($\varepsilon_1, \varepsilon_2$) of the film fitted in (a)*

We now discuss spectroscopic ellipsometry investigations performed in the spectral range of 1.1-3.0 eV. The ellipsometry spectra $\Psi$ (relative amplitude change) and $\Delta$ (relative phase shift of the polarization) for the 10 nm $Bi_2Se_3$ film taken at 70 degrees is shown in Figure 2a. This data was modeled utilizing 3 Tauc-Lorentz oscillators without invoking any oxide layer. Table 1 summarizes the fitting results. The substrate was modeled separately and then incorporated into the film+substrate model. With a mean square error value of 2.7, we could reliably extract the optical constants in addition to verifying the film thickness (~10nm). The real and imaginary parts of the complex dielectric constant are shown in Fig 2b. Two peaks are observed, the strongest one at 2.0 eV and a shoulder at 1.4 eV. The data agrees very well with reports on MBE samples [31]. Band structure calculations can identify multiple candidates responsible for these transitions (both direct and indirect) [23] [24]. Even though the spectral range of our ellipsometry measurement is limited, empirical band gaps can be inferred from the Tauc gaps ($E_g$) of the model oscillators (Table 1). Ignoring the broad background oscillator (TL1), we find that the next Tauc gap is at 0.375 eV, which is close to the gap value measured through absorption spectroscopy that is discussed next.

Table 1: Fit parameters of Tauc-Lorentz oscillators

| Oscillator | $\varepsilon_\infty = 1.44$ | | | |
| --- | --- | --- | --- | --- |
| | Amp | $E_n$ | C | $E_g$ |
| $TL_1$ | 1.239 | 1.436 | 0.375 | 0.0001 |
| $TL_2$ | 74.59 | 2.504 | 2.965 | 0.376 |
| $TL_3$ | 30.075 | 1.999 | 0.88029 | 0.5221 |

To perform the transmittance measurement, the $Bi_2Se_3$ samples were deposited on transparent quartz substrates. Two set of samples of thickness 2-10 QL were fabricated. One set of samples was capped with a few nm of amorphous BN and the other set was left uncapped. Boron nitride is a highly transparent material with a bandgap of 5.5 eV. This allowed us to protect the $Bi_2Se_3$ layer without affecting its visible and infra-red transmittance. To ascertain the impact of oxidation and other surface changes on the optical characteristics of the films, we measured the transmittance of several capped and uncapped $Bi_2Se_3$ films at various times after deposition. In Figure 3 we show the data taken at two times for the 2 QL film which is the thinnest film studied where presumably the oxidation effect, if any, should be the strongest. As expected, the uncapped sample demonstrated some change in its optical transmittance, but it is only a few percent, and mostly in the high-energy range (500-900 nm). The observed change is less than 10% even after 7 days (data not shown), proving that the optical transmittance properties were not affected to any significant degree even at the 2 QL level. The BN capped sample did not show any change as evident from the near-perfect overlap of the data taken after 6 mins and 4 hours. Taken together we infer that even though capping improves reliability, oxidation and other extrinsic effects do not dominate our optical measurements [31]. Encourage by these developments we proceeded to measure the band gap on both types of samples.

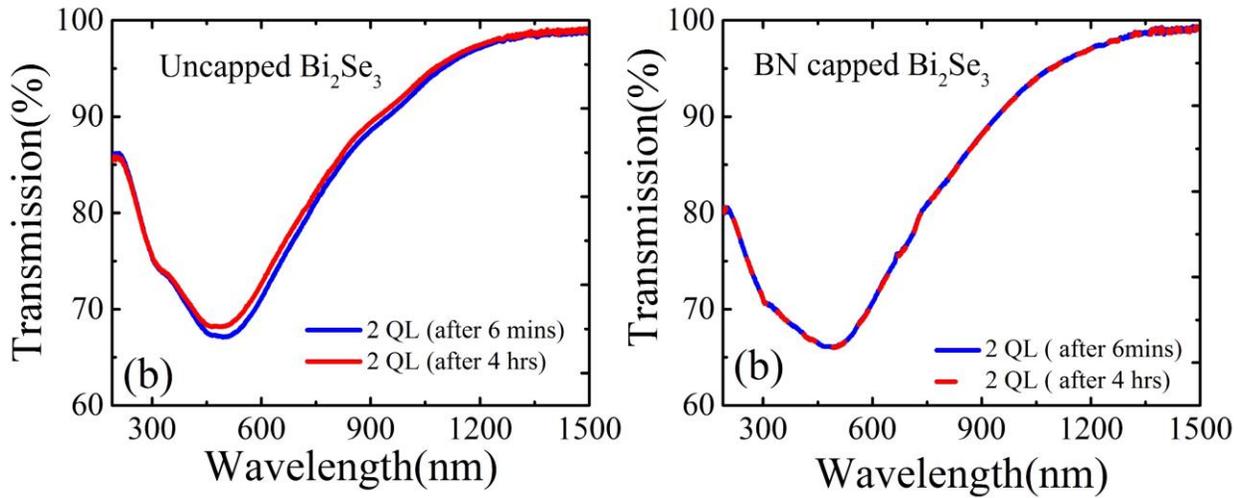

*Figure 3 Transmittance data of a 2QL uncapped (a) and BN-capped (b) $Bi_2Se_3$ film taken at different times after film deposition as indicated. The uncapped sample showed only very little variation with time.*

In figure 4 we plot the optical absorption as calculated from the transmittance data. In figure 4a, we plot the representative data for the 2 QL and 6 QL film. Various optical features are clearly seen that agree with spectroscopic ellipsometry data (Fig. 2). The most striking dissimilarity between the two films is the rigid blue-shift in the 2 QL sample compared to the 6 QL data. This clearly indicates that the fundamental band gap of these two systems are different.

The blue-shift in the absorption data continues to up to 3.5 eV, above which the optical properties overlap reasonably well.

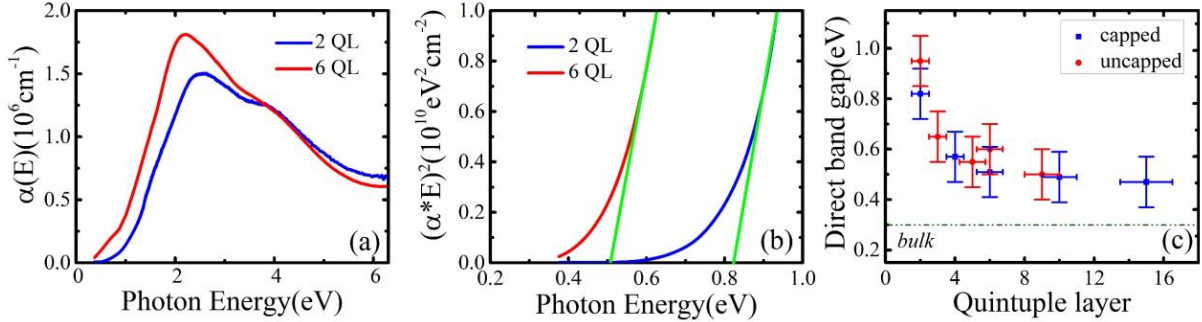

*Figure 4 (a) Optical absorption data of a 2 and 6 QL $Bi_2Se_3$ film showing a blue-shift with inverse thickness (b) Direct-gap analysis of the data shown in (a). The blue-shift is clearly quantified. (c) The direct band gap for films of different thickness. Both capped and uncapped films follow the same trend of increasing band gap with inverse thickness below 6 QL.*

To quantify the optical band gaps accurately, we plot the $(\alpha E)^2$ vs the photon energy (Fig. 4b) to reveal the direct gaps of few-layer $Bi_2Se_3$. Though this method was developed to measure optical properties of semiconductors with parabolic bands it has often been extended (successfully) to other systems. The measured band gap for the 2QL film is ~0.8 eV whereas the value for the 6QL film is ~0.5 eV. In Fig. 4c, we plot the band gaps of all the few-layer films (both capped and uncapped). The strong increase in band gap with decreasing thickness is seen in both types of samples. The obtained band gap for the 15 QL film is close to bulk value (~0.3 eV), that demonstrates that the systematic error in the band gap estimation is small, if any (less than 0.1 eV).

Our observations are consistent with strong differences in the electronic character of two-dimensional $Bi_2Se_3$ compared to bulk. Previous efforts have highlighted that finite size effect is particularly important when the top and bottom surfaces starts to interact with each other [25, 26], the onset of which starts below six QL. It is therefore not a coincidence that the most significant band gap change is for thicknesses below six quintuple layers. We therefore hypothesize while optical transmittance measurements might not be directly sensitive *only* to the surface states, we are indirectly measuring the effects of it. It is also worth noting that Vargas *et al.* reported a theoretical band gap of ~0.8 eV for a 2QL $Bi_2Se_3$ [29], which is also in excellent quantitative agreement with our results. Therefore, we conjecture that quantum confinement effects might already play a significant role at the 2D level. Additional lateral confinement will only amplify this effect. At present, we have not confirmed this effect for a single $Bi_2Se_3$ layer but believe that it will be a robust effect. Our data sheds more evidence into the scaling behavior of layered systems that can potentially have interesting consequences in future nanoelectronics devices.

In conclusion, we have provided optical evidence of a rigid blue shift in $Bi_2Se_3$ thin films as we approach the two-dimensional limit. High-quality, oriented, few-layer $Bi_2Se_3$ films was grown using magnetron sputtering and their structural and optical properties was investigated using X-ray, Raman, and spectroscopic ellipsometry, and transmittance spectroscopy. Up to 0.5 eV change in band gap is observed (compared the bulk), and most significantly below 6 QL. The effect is robust and is observed in both capped and uncapped films. Our observations are consistent with reports that show dramatic changes in electronic behavior of topological materials as they approach lower dimensions. Our work highlights that topological materials follow a complex scaling behavior that requires significant investigation before serious application of such materials for nanoelectronics devices are considered.

## ACKNLOWLEDGEMENTS

DM would like to acknowledge support from startup funds from Southern Illinois University (SIU), SIU Elevation research grant, and SIU Materials Technology Center. Raman measurements shown here was carried out at the Frederick Seitz Materials Research Laboratory Central Research Facilities, University of Illinois.